\input phyzzx
\def\psibar{\overline\psi}
\def\phibar{\overline\phi}
\def\Dslash{D\kern-0.15em\raise0.17ex\llap{/}\kern0.15em\relax}
\def\Dslashl{\mathop{\Dslash}\limits^\leftarrow}
\def\Dmul{\mathop{D_\mu}\limits^\leftarrow}
\def\partialmul{\mathop{\partial_\mu}\limits^\leftarrow}
\def\Rl{\mathop{R}\limits^\leftarrow}
\def\bl{\mathop{b}\limits^\leftarrow}
\def\sslash{s\kern-0.026em\raise0.17ex\llap{/}%
          \kern0.026em\relax}
\def\kslash{k\kern-0.026em\raise0.17ex\llap{/}%
          \kern0.026em\relax}
\def\pslash{p\kern-0.026em\raise0.17ex\llap{/}%
          \kern0.026em\relax}
\def\qslash{q\kern-0.026em\raise0.17ex\llap{/}%
          \kern0.026em\relax}
%
\REF\PAU{
W. Pauli and F. Villars,
Rev.\ Mod.\ Phys.\ 21 (1949) 434;\hfill\break
S. N. Gupta,
Proc.\ Phys.\ Soc.\ A66 (1953) 129.}
\REF\SHA{
For recent review, see,
Y. Shamir, {\it in\/} Lattice 1995, hep-lat/9509023;\hfill\break
H. Neuberger, hep-lat/9511001.}
\REF\WIL{
K. G. Wilson, {\it in\/} New phenomena in subnuclear physics, ed.\
by A. Zichichi (Plenum Press, New York, 1977).}
\REF\NIE{
H. B. Nielsen and M. Ninomiya,
Nucl.\ Phys.\ B185 (1981) 20;\hfill\break
L. H. Karsten,
Phys.\ Lett.\ B104 (1981) 315.}
\REF\FRO{
S. A. Frolov and A. A. Slavnov,
Nucl.\ Phys.\ B411 (1994) 647.}
\REF\SLA{
A. A. Slavnov,
Phys.\ Lett.\ B319 (1993) 231; B348 (1995) 553.}
\REF\FROL{
S. A. Frolov and A. A. Slavnov,
Phys.\ Lett.\ B309 (1993) 344.}
\REF\NAR{
R. Narayanan and H. Neuberger,
Phys.\ Lett.\ B302 (1993) 62;\hfil\break
S. Aoki and Y. Kikukawa,
Mod.\ Phys.\ Lett.\ A8 (1993) 3517;\hfil\break
K. Fujikawa,
Nucl.\ Phys.\ B428 (1994) 169;
{\it in\/} Festschrift in honor of H.~Banerjee,
hep-th/9506003, to appear in Indian Jour.\ of Phys.,\hfill\break
K. Okuyama and H. Suzuki,
hep-th/9603062; Phys.\ Lett.\ B382 (1996) 117;\hfill\break
L. N. Chang and C. Soo, hep-th/9606174.}
\REF\KAR{
L. H. Karsten and J. Smit,
Nucl.\ Phys.\ B144 (1978) 536.}
\REF\KER{
W. Kerler, Phys.\ Rev.\ D23 (1981) 2384.}
\REF\SEI{
E. Seiler and I. Stamatescu, Phys.\ Rev.\ D25 (1982) 2177.}
\REF\FUJ{
K. Fujikawa,
Z.\ Phys.\ C25 (1984) 179.}
\REF\FUJI{
K. Fujikawa,
Phys.\ Rev.\ Lett.\ 42 (1979) 1195;
Phys.\ Rev.\ D21 (1980) 2848; D22 (1980) 1499(E).}
\REF\THO{
G. 't Hooft,
Phys.\ Rev.\ Lett.\ 37 (1976) 8;
Phys.\ Rev.\ D14 (1976) 3432; D18 (1978) 2199(E).}
%
\baselineskip=7mm
\overfullrule=0pt
\pubnum={
hep-th/9608025}
\date={December 1996}
\titlepage
\title{Remark on Pauli--Villars Lagrangian on the Lattice}
\author{%
Kazunobu Haga, Hiroshi Igarashi, Kiyoshi Okuyama\break
and\break
Hiroshi Suzuki\foot{%
e-mail: hsuzuki@mito.ipc.ibaraki.ac.jp}}
\address{%
Department of Physics, Ibaraki University, Mito 310, Japan}
\abstract{%
It is interesting to superimpose the Pauli--Villars regularization
on the lattice regularization. We illustrate how this scheme
works by evaluating the axial anomaly in a simple lattice fermion
model, the Pauli--Villars Lagrangian with a gauge non-invariant
Wilson term. The gauge non-invariance of the axial anomaly, caused
by the Wilson term, is remedied by a compensation among
Pauli--Villars regulators in the continuum limit. A subtlety in
Frolov--Slavnov's scheme for an {\it odd\/} number of chiral fermions
in an anomaly free complex gauge representation, which requires an
infinite number of regulators, is briefly mentioned.
}
\endpage

It seems interesting to put a Pauli--Villars type Lagrangian level
regularization on the lattice. The interest is twofold:
The Pauli--Villars regularization~[\PAU] for fermion one-loop
diagrams can be expressed as a Lagrangian of regulators
(bosonic and fermionic spinors). In actual perturbative
calculations however, the Lagrangian has to be supplemented with
additional prescriptions, such that the momentum of propagators have
to be assigned in the same way for all the fields, and the integrand
in the momentum integral has to be summed before the integration.
Once the Lagrangian is put on the lattice, no prescription are needed
and one is free to choose any momentum assignment. (To get a finite
gauge invariant result in the continuum limit one has to assign the
momenta of all fields on the lattice in the same way.)

More interestingly and importantly, ``superimposing''
a different kind of regularization on the lattice regularization
may give some clue to the lattice regularization of chiral gauge
theories. No manifestly gauge invariant lattice formulation of the
chiral gauge theory, being consistent with the unitarity and the
locality, is yet known~[\SHA]. In particular, for a chiral fermion in
a complex gauge representation, it is impossible to introduce in a
gauge invariant way the Wilson term~[\WIL] to eliminate unwanted
species doublers.\foot{%
For chiral fermions in a real-positive gauge representation, and
for even number of chiral fermions in a pseudoreal representation,
it is possible to introduce a gauge invariant (Majorana-type) Wilson
term.} The difficulty of a manifestly gauge invariant lattice
formulation of chiral gauge theories is highlighted by the No-Go
theorem~[\NIE].

The basic idea of ``superimposing'' is quite simple. Let us consider,
for example, the naive momentum cutoff regularization applied to
fermion one-loop diagrams in QED. This regularization breaks the
gauge invariance, generating gauge non-invariant contributions.
However we may use in addition say, the gauge invariant dimensional
regularization. With this superimposed regularization, the infinite
momentum cutoff limit can be taken and we are left with gauge
invariant expressions in the dimensional regularization. Of course
there is no real need to break the gauge invariance by
introducing the momentum cutoff in this example, but with the lattice
regularization, it is not obvious how to treat chiral
fermions in a manifestly gauge invariant manner. To perform this
program it is clearly crucial that there exists a regularization which
preserves the gauge symmetry and simultaneously is congenial to the
lattice regularization.

In fact a proposal based on this idea has been made by Frolov and
Slavnov~[\FRO] (see also~[\SLA]). They used the gauge invariant
generalized Pauli--Villars regularization~[\FROL,\NAR] for chiral
fermions in an anomaly free complex representation, and discussed
that taking the continuum limit $a\to0$ ($a$ is the lattice spacing)
with an appropriately scaled regulator mass~$M(a)\ll1/a$, the regulator
fields compensate the effect of gauge non-invariant Wilson term and that
the gauge invariant regularized continuum theory~[\FROL] is reproduced. 

We illustrate in this letter how this scheme works by evaluating the
axial anomaly [\KAR,\KER,\SEI] in a simpler lattice fermion model,
the Pauli--Villars Lagrangian with a gauge {\it non}-invariant
Wilson term in the lattice {\it vector\/} gauge theory. The relation
to the proposal in~[\FRO] and the possible implication will be
commented on later.

Before considering the Pauli--Villars regulators, let us study
for a while a massive Dirac fermion coupled to a background gauge
field and the axial U(1) Ward identity. The naive fermion action is
$$
   I\equiv\sum_x\psibar(x)\bigl[i\Dslash(x)-m\bigr]\psi(x)
   =\sum_x\psibar(x)\bigl[-i\Dslashl(x)-m\bigr]\psi(x),
\eqn\one
$$
where the covariant derivative on the lattice $D_\mu(x)$ has been
defined by
$$
\eqalign{
   &D_\mu(x)\equiv
   {1\over2a}\Bigl[U_\mu(x)e^{a\partial_\mu}
                   -e^{-a\partial_\mu}U_\mu^\dagger(x)\Bigr],
\cr
   &\Dmul(x)\equiv
   -{1\over2a}\Bigl[U_\mu(x)e^{-a\partialmul}
                   -e^{a\partialmul}U_\mu^\dagger(x)\Bigr].
\cr
}
\eqn\two
$$
In the continuum limit $a\to0$, we parameterize the link variable as
$U_\mu(x)=e^{iagA_\mu(x)}$.
As is well known, the naive action \one\ contains unwanted
species doublers. Therefore we add the Wilson term to decouple them in
the continuum limit, but an artificially chosen gauge
{\it non}-invariant one:
$$
   I_W\equiv\sum_x\psibar(x)R(x)\psi(x)
      =\sum_x\psibar(x)(-\Rl(x))\psi(x),
\eqn\three
$$
with
$$
   R(x)\equiv{r\over2a}\sum_\mu\Bigl(e^{a\partial_\mu}
                   +e^{-a\partial_\mu}-2\Bigr),\quad
   \Rl(x)\equiv-{r\over2a}\sum_\mu\Bigl(e^{-a\partialmul}
                   +e^{a\partialmul}-2\Bigr).
\eqn\four
$$
Although $I_W$ is irrelevant in the naive continuum limit,
the effect of hard breaking of chiral and gauge
symmetries survives in the axial anomaly as we will see below.

The axial U(1) Ward identity for the lattice action $I+I_W$ is
derived by performing a change of variable
$\psi(x)\to e^{i\alpha(x)\gamma_5}\psi(x)$ and
$\psibar(x)\to\psibar(x)e^{i\alpha(x)\gamma_5}$ in the partition
function. For an infinitesimal $\alpha(x)$, the action changes as
$$
\eqalign{
   &I\to I+\sum_x\alpha(x)
   \Bigl[\partial_\mu J_5^\mu(x)
         -2im\psibar(x)\gamma_5\psi(x)\Bigr],
\cr
   &I_W\to I_W-\sum_x\alpha(x)B(x),
\cr
}
\eqn\five
$$
where the divergence of the axial U(1) current has been defined
by
$$
\eqalign{
   \partial_\mu J_5^\mu(x)&\equiv
   \psibar(x)\Dslash(x)\gamma_5\psi(x)
   +\psibar(x)\Dslashl(x)\gamma_5\psi(x)
\cr
   &=\sum_\mu\partial_\mu\Bigl[
    \psibar(x)\gamma^\mu\gamma_5\psi(x)\Bigr]+O(a),
\cr
}
\eqn\six
$$
(the second line is the naive continuum limit) and
$$
   B(x)\equiv
   -\psibar(x)i\gamma_5R(x)\psi(x)
   +\psibar(x)i\gamma_5\Rl(x)\psi(x),
\eqn\seven
$$
the explicit axial U(1) breaking part. Since the functional
integration measure is invariant under the change of variable with
the lattice regularization, the Ward identity reads
$$
   \VEV{\partial_\mu J_5^\mu(x)}=
   \VEV{2im\psibar(x)\gamma_5\psi(x)}+\VEV{B(x)}.
\eqn\eight
$$

Let us evaluate the right hand side of \eight. We first concentrate
on the vacuum expectation value of $B(x)$,\foot{%
Our calculation method is similar to that of~[\SEI], but seems
rather simpler.}
$$
\eqalign{
   \VEV{B(x)}&=\tr i\gamma_5R(x)\VEV{\psi(x)\psibar(y)}\bigr|_{x=y}
   -\tr\VEV{\psi(x)\psibar(y)}i\gamma_5\Rl(y)\bigr|_{x=y}
\cr
   &\equiv b(x)+\bl(x).
\cr
}
\eqn\nine
$$
They are evaluated by the lattice propagator in the presence of
background gauge field,
$$
\eqalign{
   &b(x)\equiv
   -\tr i\gamma_5R(x){1\over i\Dslash(x)-m+R(x)}\delta(x,y)
   \bigr|_{x=y},
\cr
   &\bl(x)\equiv
   -\tr\delta(x,y){1\over i\Dslashl(y)+m+\Rl(y)}
   i\gamma_5\Rl(y)\bigr|_{x=y},
\cr
}
\eqn\ten
$$
where the delta function on the lattice is defined by
$\delta(x,y)\equiv\delta_{x,y}/a^4=
\int_{-\pi/a}^{\pi/a}d^4k\*e^{ik(x-y)}/(2\pi)^4$ and hence
$$
\eqalign{
   b(x)&=-\tr{1\over a^4}\int_{-\pi}^\pi{d^4k\over(2\pi)^4}
    \,e^{-ikx/a}i\gamma_5R(-i\Dslash-m+R)
\cr
   &\quad\times
    {1\over
    -\sum_\mu D_\mu^2+(m-R)^2
    +\sum_{\mu,\nu}[\gamma^\mu,\gamma^\nu][D_\mu,D_\nu]/4
    +i[\Dslash,R]}\,e^{ikx/a}.
\cr
}
\eqn\eleven
$$
In deriving the above expression, we have multiplied
$(-i\Dslash-m+R)$ on the numerator and on the denominator, and
used a relation
$\Dslash^2=-\sum_\mu D_\mu^2
+\sum_{\mu,\nu}[\gamma^\mu,\gamma^\nu]\*[D_\mu,D_\nu]/4$.
Next noting
$$
   e^{-ikx/a}D_\mu e^{ikx/a}={i\over a}\sin k_\mu+\widetilde D_\mu,
   \quad
   e^{-ikx/a}Re^{ikx/a}={r\over a}\sum_\mu(\cos k_\mu-1)
   +\widetilde R,
\eqn\twelve
$$
where
$$
\eqalign{
   \widetilde D_\mu&\equiv
   {1\over2a}\left[e^{ik_\mu}(U_\mu e^{a\partial_\mu}-1)
    +e^{-ik_\mu}(1-e^{-a\partial_\mu}U_\mu^\dagger)\right]
\cr
   &=\cos k_\mu(\partial_\mu+igA_\mu)+O(a),
\cr
}
\eqn\thirteen
$$
and
$$
\eqalign{
   \widetilde R&\equiv
   {r\over2a}\sum_\mu\left[e^{ik_\mu}(e^{a\partial_\mu}-1)
    -e^{-ik_\mu}(1-e^{-a\partial_\mu})\right]
\cr
   &=ir\sum_\mu\sin k_\mu\partial_\mu+O(a),
\cr
}
\eqn\fourteen
$$
we have
$$
\eqalign{
   &b(x)
\cr
   &=-\tr{1\over a^4}\int_{-\pi}^\pi{d^4k\over(2\pi)^4}\,
    i\gamma_5\Bigl[r\sum_\mu(c_\mu-1)+a\widetilde R\Bigr]
   \Bigl[\sslash+r\sum_\nu(c_\nu-1)
         -ia\widetilde\Dslash-am+a\widetilde R\Bigr]
\cr
   &\quad\times
   \biggl\{
    -\sum_\rho(is_\rho+a\widetilde D_\rho)^2
    +\Bigl[r\sum_\rho(c_\rho-1)-am+a\widetilde R\Bigr]^2
\cr
   &\qquad\qquad\qquad\qquad\qquad\qquad\qquad\quad
   +{a^2\over4}\sum_{\rho,\sigma}[\gamma^\rho,\gamma^\sigma]
     [\widetilde D_\rho,\widetilde D_\sigma]
    +ia^2[\widetilde\Dslash,\widetilde R]\biggr\}^{-1}\cdot1,
\cr
}
\eqn\fifteen
$$
where the trigonometric functions have been abbreviated as
$s_\mu\equiv\sin k_\mu$ and $c_\mu\equiv\cos k_\mu$.

In \fifteen, the expansion with respect to $a$ is straightforward
because the trace of gamma matrices requires at least four of them
($\tr\gamma_5\gamma^\mu\gamma^\nu\gamma^\rho\gamma^\sigma
=-4\varepsilon^{\mu\nu\rho\sigma}$). Finally using
$$
\eqalign{
   &[\widetilde D_\mu,\widetilde D_\nu]=
   igc_\mu c_\nu F_{\mu\nu}+O(a),
\cr
   &[\widetilde D_\mu,\widetilde R]=
   rg\sum_\nu c_\mu s_\nu(\partial_\nu A_\mu)+O(a),
\cr
}
\eqn\sixteen
$$
where
$F_{\mu\nu}\equiv\partial_\mu A_\nu-\partial_\nu A_\mu
+ig[A_\mu,A_\nu]$,
$$
   \lim_{a\to0}b(x)=
   -{ig^2\over(2\pi)^4}I_1(r)\varepsilon^{\mu\nu\rho\sigma}
        \tr F_{\mu\nu}F_{\rho\sigma}
   -{ig^2\over(2\pi)^4}I_2(r)\varepsilon^{\mu\nu\rho\sigma}
        \tr\partial_\mu A_\nu F_{\rho\sigma}.
\eqn\seventeen
$$
In \seventeen, $I_1(r)$ and $I_2(r)$ are the well-known lattice
integrals~[\KAR,\SEI]:
$$
\eqalign{
   &I_1(r)=r^2\int_{-\pi}^\pi d^4k\,
   {\displaystyle
    \prod_\mu c_\mu\Bigl[\sum_\nu(1-c_\nu)\Bigr]^2\over
    \displaystyle
    \biggl\{\sum_\rho s_\rho^2
           +r^2\Bigl[\sum_\rho(1-c_\rho)\Bigr]^2\biggr\}^3},
\cr
   &I_2(r)=-r^2\int_{-\pi}^\pi d^4k\,
   {\displaystyle
   \sum_\mu s_\mu^2\prod_{\nu\neq\mu}c_\nu
   \sum_\sigma(1-c_\sigma)\over
   \displaystyle
   \biggl\{\sum_\rho s_\rho^2+r^2\Bigl[\sum_\rho(1-c_\rho)\Bigr]^2
           \biggr\}^3},
\cr
}
\eqn\eighteen
$$
and obey $I_1(r)+I_2(r)=-\pi^2/2$.
A similar calculation shows $\bl(x)=b(x)$. Thus finally we arrive at
$$
\eqalign{
   \lim_{a\to0}\VEV{\partial_\mu J_5^\mu(x)}&=
   \lim_{a\to0}\VEV{2im\psibar(x)\gamma_5\psi(x)}
   +{ig^2\over16\pi^2}\varepsilon^{\mu\nu\rho\sigma}
    \tr F_{\mu\nu}F_{\rho\sigma}
\cr
   &\quad+{ig^2\over4\pi^2}I_2(r)
   \varepsilon^{\mu\nu\rho\sigma}
    \tr\bigl[\partial_\mu(A_\nu\partial_\rho A_\sigma
                                  +igA_\nu A_\rho A_\sigma)\bigr].
\cr
}
\eqn\nineteen
$$

Several comments are in order: It is well-known that, with the gauge
invariant Wilson term~[\WIL], the species doublers effectively act as
the Pauli--Villars regulators (i.e., with an
alternative axial charge~[\KAR])
and the gauge invariant correct axial anomaly is reproduced in the
continuum limit~[\KAR,\SEI]. In our present case, the last term on
the right hand side of \nineteen\ is not gauge invariant (the first
term will be shown to be gauge invariant). It would be gauge invariant
if the last coefficient was $2/3$ instead of $1$. It also depends on
the Wilson parameter $r$, although for an infinitesimal $r$ the term
vanishes $\lim_{r\to0}I_2(r)=0$. Therefore the effect of hard
breaking of chiral and gauge symmetries in the Wilson term
\three\ survives in the continuum limit. In other words, our
identification of the axial current \six\ was not gauge invariant
(with the gauge non-invariant Wilson term \three), and the operator
does not coincide with the naive continuum limit. We may redefine the
axial current as only the first line in \nineteen\ survives and to
restore the gauge invariance. Such an intricacy of an operator
identification in the lattice regularization in the view point of the
anomaly and the usefulness of superimposing a different regularization
are emphasized in~[\FUJ].

For our purpose, the first term in \nineteen\ is important:
$$
\eqalign{
   &\VEV{2im\psibar(x)\gamma_5\psi(x)}
\cr
   &=2i\tr{1\over a^4}\int_{-\pi}^\pi{d^4k\over(2\pi)^4}\,
   am\gamma_5
   \Bigl[\sslash+r\sum_\nu(c_\nu-1)
         -ia\widetilde\Dslash-am+a\widetilde R\Bigr]
\cr
   &\quad\times
   \biggl\{
    -\sum_\rho(is_\rho+a\widetilde D_\rho)^2
    +\Bigl[r\sum_\rho(c_\rho-1)-am+a\widetilde R\Bigr]^2
\cr
   &\qquad\qquad\qquad\qquad\qquad\qquad\qquad\quad
   +{a^2\over4}\sum_{\rho,\sigma}[\gamma^\rho,\gamma^\sigma]
     [\widetilde D_\rho,\widetilde D_\sigma]
    +ia^2[\widetilde\Dslash,\widetilde R]\biggr\}^{-1}\cdot1.
\cr
}
\eqn\twenty
$$
A simple expansion by $a$ is impossible in \twenty\ because of the
singular infrared behavior near $k\sim0$. Thus we divide the
integration region to the ``outer'' region $|k|>\delta$, which is
free of the infrared divergence, and the ``inner'' region
$|k|\leq\delta$ with an infinitesimal $\delta\ll1$~[\KAR].
In the outer region, we may safely expand the integrand with respect
to $a$, yielding
$$
   \lim_{a\to0}\VEV{2im\psibar(x)\gamma_5\psi(x)}_{\rm outer}=0.
\eqn\twentyone
$$
In the inner region we expand it by the external gauge field:
$$
\eqalign{
   &\VEV{2im\psibar(x)\gamma_5\psi(x)}_{\rm inner}=2im\Gamma_5
\cr
   &+2im\sum_{n=1}^\infty{1\over n!}\prod_{i=1}^n
   \biggl[\sum_{x_i,\mu_i}A_{\mu_i}(x_i)\int{d^4p_i\over(2\pi)^4}\,
   e^{ip_i(x-x_i)}e^{-iap_i/2}\biggr]
   \Gamma_5^{\mu_1\mu_2\cdots\mu_n}(p_1,p_2,\cdots,p_n).
\cr
}
\eqn\twentytwo
$$
For example, $\Gamma_5=\Gamma_5^\mu(p)=0$, and
$$
\eqalign{
   &\Gamma_5^{\mu\nu}(p,q)
\cr
   &=-\tr\gamma_5\int_{-\delta/a}^{\delta/a}
   {d^4k\over(2\pi)^4}
   \Bigl[S(k+p+q)V^\mu(k+p+q,k+q)S(k+q)V^\nu(k+q,k)S(k)
\cr
   &\qquad\qquad\qquad\qquad\qquad
   +(\mu\leftrightarrow\nu,p\leftrightarrow q)\Bigr],
\cr
}
\eqn\twentythree
$$
where\foot{%
The so-called anomalous vertex~[\KAR] does not contribute to this
function.}
$$
\eqalign{
   &S(k)\equiv\Bigl[\sum_\mu\gamma^\mu{1\over a}\sin ak_\mu+m
         +{r\over a}\sum_\mu(1-\cos ak_\mu)\Bigr]^{-1},
\cr
   &V^\mu(k_1,k_2)\equiv
   g\biggl[
   \gamma^\mu\cos a\Bigl({1\over2}{k_1}_\mu+{1\over2}{k_2}_\mu\Bigr)
           +r\sin a\Bigl({1\over2}{k_1}_\mu+{1\over2}{k_2}_\mu\Bigr)
   \biggr].
\cr
}
\eqn\twentyfour
$$
Since what to be worried about is the infrared divergence, we may
simply expand the numerator by $a$ as
$V^\mu(k+p+q,k+q)=g\gamma^\mu+O(a)$ etc.
In the denominator, we may expand the propagator as,
$S(k+p)=[\sum_\mu(k_\mu+p_\mu)+m+O(\delta^2)]^{-1}$ because
$\delta\ll1$. Note that the subleading contributions from the
denominator always give a ultraviolet convergent integral. From
these arguments, we have
$$
\eqalign{
   &\lim_{a\to0}\Gamma_5^{\mu\nu}(p,q)
\cr
   &=-g^2\tr\gamma_5\lim_{a\to0}\int_{-\delta/a}^{\delta/a}
   {d^4k\over(2\pi)^4}\,
\cr
   &\qquad\times\left[{1\over\kslash+\pslash+\qslash+m}\gamma^\mu
   {1\over\kslash+\qslash+m}\gamma^\nu
   {1\over\kslash+m}+
   (\mu\leftrightarrow\nu,p\leftrightarrow q)\right]+O(\delta^2)
\cr
   &=-{g^2\over4\pi^2}m\sum_{\rho,\sigma}
   \varepsilon^{\mu\nu\rho\sigma}p_\rho q_\sigma
   \int_0^12ydy\int_0^1dx
\cr
   &\qquad\qquad\quad\times
   {1\over m^2-y(1-y)q^2-2xy(1-y)p\cdot q-xy(1-xy)p^2}+O(\delta^2).
\cr
}
\eqn\twentyfive
$$
After the safe limit $\delta\to0$, \twentyfive\ is nothing but the
expression in the continuum theory. The same consideration can be
repeated for higher point functions, and we may summarize this
fact compactly as
$$
   \lim_{a\to0}\VEV{2im\psibar(x)\gamma_5\psi(x)}
   =2i\lim_{y\to x}\tr
    m\gamma_5{1\over i\Dslash_c-m}\delta(x-y),
\eqn\twentysix
$$
where $\Dslash_c\equiv\gamma^\mu(\partial_\mu+igA_\mu)$
is the covariant derivative in the {\it continuum\/} theory.
Note \twentysix\ is {\it finite\/} without any further regularization
as the last line of \twentyfive\ shows.
The first term in \nineteen\ is therefore gauge invariant.
As we have seen in \twentyone, only the physical fermion near
$k\sim0$ contributes to the operator \twenty, and the effect of
Wilson term which couples only to the species doublers, is invisible
in the continuum limit.

Combining \nineteen\ and \twentysix, we have for a single massive
Dirac fermion,
$$
\eqalign{
   &\lim_{a\to0}\VEV{\partial_\mu J_5^\mu(x)}
\cr
   &=
   2i\lim_{y\to x}\tr
    m\gamma_5{1\over i\Dslash_c-m}\delta(x-y)
\cr
   &\quad+{ig^2\over16\pi^2}\varepsilon^{\mu\nu\rho\sigma}
    \tr F_{\mu\nu}F_{\rho\sigma}
   +{ig^2\over4\pi^2}I_2(r)
   \varepsilon^{\mu\nu\rho\sigma}
    \tr\bigl[\partial_\mu(A_\nu\partial_\rho A_\sigma
                                  +igA_\nu A_\rho A_\sigma)\bigr].
\cr
}
\eqn\twentyseven
$$

Let us now introduce the Pauli--Villars Lagrangian on the lattice.
We introduce the regulator fields $\psi_n$, where $n=1,2,\cdots,N$,
and $\psi_0$ the original fermion to be regularized. We assign
the even number index for fermionic fields and the odd index
for bosonic ones. We also denote the mass of those fields as $m_n$
and assume the masses of the regulator fields are of the order of the
``cutoff'' parameter $\Lambda$.
The Pauli--Villars regularization condition~[\PAU] requires
$\sum_{n=0}^N(-1)^n=\sum_{n=0}^N(-1)m_n^2=0$. For the combined
system, we can use \twentyseven\ for each fermionic field and
that with a reversed sign for each bosonic field. For example,
by summing up the first term, we have
$$
\eqalign{
   &2i\lim_{y\to x}\tr
   \sum_{n=0}^N(-1)^nm_n\gamma_5{1\over i\Dslash_c-m_n}\delta(x-y)
\cr
   &=\VEV{2im_0\psibar_0\gamma_5\psi_0}
   +2i\lim_{y\to x}\tr\gamma_5f(\Dslash_c^2/\Lambda^2)\delta(x-y),
\cr
}
\eqn\twentyeight
$$
where we have defined the regulator function
$$
   f(t)\equiv
   -\sum_{n=1}^N{(-1)^nm_n^2/\Lambda^2\over t+m_n^2/\Lambda^2}.
\eqn\twentynine
$$
It follows from the definition and the Pauli--Villars condition,
$f(0)=1$ and $f(t)=m_0^2/(\Lambda^2 t)+O(1/t^2)$. For example,
a possible choice is
$m_1^2=\Lambda^2$, $m_2^2=2\Lambda^2$, $m_3^2=\Lambda^2+m_0^2$,
and $\lim_{\Lambda\to\infty}f(t)=2/(t+1)(t+2)$.

On the other hand, the second line in \twentyseven\ cancels out
among the fermionic and the bosonic fields because it
is independent of the mass of the field. Therefore the total axial
anomaly is given by
$$
   \lim_{\Lambda\to\infty}\lim_{a\to0}
   \VEV{\partial_\mu J_5^\mu(x)}
   =\VEV{2im_0\psibar_0\gamma_5\psi_0}
   +{ig^2\over16\pi^2}\varepsilon^{\mu\nu\rho\sigma}
    \tr F_{\mu\nu}F_{\rho\sigma}.
\eqn\thirty
$$
In deriving this, we have used the fact $f(0)=1$ and
$f(\infty)=f'(\infty)=\cdots=0$ for $\Lambda\to\infty$ and the actual
calculation is identical to that of~[\FUJI]. We have thus recovered the
correct gauge invariant form of the axial anomaly.
The present analysis may be repeated for the conformal
anomaly of the Wilson fermion, for which the Wilson term gives rise
$-15$ times the correct coefficient~[\FUJ]. We expect that the
correct coefficient will be reproduced in the present scheme, because
the {\it real\/} Pauli--Villars regulators, i.e., with an alternative
{\it statistics\/}, always eliminate the effect of Wilson term.

Finally we briefly comment on the implication of above
demonstration for the proposal in~[\FRO]. By a suitable change of
variable,
$$
\eqalign{
   &\chi(x)={1\over\sqrt{2}}\left[
   P_R\psi(x)+P_LC\Gamma_{11}C_D\psibar^T(x)\right],
\cr
   &\varphi(x)={1\over\sqrt{2}}\left[
   P_R\phi(x)+P_LCC_D\phibar^T(x)\right],
\cr
   &\widetilde\varphi(x)={1\over\sqrt{2}}\left[
   P_R\widetilde\phi(x)+P_LCC_D\overline{\widetilde\phi}^T(x)\right],
\cr
}
\eqn\thirtyone
$$
($\chi(x)$ is a fermionic field and $\varphi(x)$,
$\widetilde\varphi(x)$ are bosonic fields), it is possible to rewrite
the lattice Pauli--Villars Lagrangian of~[\FRO] {\it basically\/} in
the form of $I+I_W$ in \one\ and \three. Note however that the number
of the degree of freedom is doubled in \thirtyone\ because, among four
chirality components of $\chi(x)$ and $\overline\chi(x)$, only two of
them are independent ($P_R\psi(x)$ and $\psibar(x)P_L$ are independent
in the original chiral model). Therefore the above result
\twentyseven\ {\it divided by two\/} with $m=0$ might be regarded as
the fermion number U(1) anomaly~[\THO] of the original massless chiral
fermion in the lattice model of~[\FRO].

>From the analyses in [\FROL,\NAR], we know that it is possible to
construct a gauge invariant Pauli--Villars Lagrangian, by
utilizing a {\it finite\/} number of regulator fields, for chiral
fermions in a real-positive representation, and for an {\it even}
number of chiral fermions in a pseudoreal or an anomaly free complex
representation. In these cases the situation would be the same as
the above analysis and we would have \thirty\ with $m_0=0$ as the
twice of the fermion number anomaly, i.e., the correct result,
$\lim_{\Lambda\to\infty}\lim_{a\to0}\VEV{\partial_\mu J^\mu(x)}
=ig^2\varepsilon^{\mu\nu\rho\sigma}\tr F_{\mu\nu}F_{\rho\sigma}/
(32\pi^2)$.

For an {\it odd\/} number of chiral fermions in a pseudoreal
representation and in an anomaly free complex representation, it is
necessary to introduce an {\it infinite\/} number of
regulators~[\FROL,\NAR]. For the former case, it is possible to
first put a finite number of them on the lattice, in a way that
$\sum_{n=0}^N(-1)^n=0$, and then to take the limit $N\to\infty$.
Thus the above result might hold even in the $N\to\infty$
limit. On the other hand, the situation seems more subtle for the
latter case, because the condition $\sum_{n=0}^N(-1)^n=0$ is
{\it never\/} satisfied for a finite $N$~[\FROL,\FRO]. The reason is
that, while the original chiral fermion belongs to a complex
representation, all the regulator fields belong to a ``doubled''
representation~[\FROL,\FRO,\NAR], namely the contribution of one
regulator field is twice of the original fermion. For example, if we
put the same finite number of fermionic and bosonic regulators first,
and then take the limit $N\to\infty$, it would correspond to
$\lim_{N\to\infty}\sum_{n=0}^N(-1)^n=1$ in the above notation.
On the other hand, the regulator function \twentynine\ would be given
by $\lim_{N\to\infty}f(t)=\pi\sqrt{t}/\sinh(\pi\sqrt{t})$~[\FRO].
Therefore the fermion number anomaly would be given by
$$
\eqalign{
   &\lim_{\Lambda\to\infty}\lim_{N\to\infty}\lim_{a\to0}
   \VEV{\partial_\mu J^\mu(x)}=
   {ig^2\over32\pi^2}\varepsilon^{\mu\nu\rho\sigma}
    \tr F_{\mu\nu}F_{\rho\sigma}
\cr
   &\quad+{ig^2\over32\pi^2}\varepsilon^{\mu\nu\rho\sigma}
    \tr F_{\mu\nu}F_{\rho\sigma}
   +{ig^2\over8\pi^2}I_2(r)
   \varepsilon^{\mu\nu\rho\sigma}
    \tr\bigl[\partial_\mu(A_\nu\partial_\rho A_\sigma
                                  +igA_\nu A_\rho A_\sigma)\bigr],
\cr
}
\eqn\thirtytwo
$$
and the gauge non-invariant piece survives.
The conclusion in~[\FRO], on the other hand, would imply
$$
   \lim_{\Lambda\to\infty}\lim_{a\to0}\lim_{N\to\infty}
   \VEV{\partial_\mu J^\mu(x)}
   ={ig^2\over32\pi^2}\varepsilon^{\mu\nu\rho\sigma}
    \tr F_{\mu\nu}F_{\rho\sigma}.
\eqn\thirtythree
$$

The above (admittedly handwaving) argument, \thirtytwo\ and
\thirtythree, shows that the anomaly is quite sensitive on
detailed way of the limit $a\to0$ and $N\to\infty$, in
Frolov--Slavnov's scheme for an {\it odd\/} number of chiral fermions
in an anomaly free complex representation.

In conclusion, we have illustrated how the superimposing of
the Pauli--Villars regularization on the lattice regularization
works, utilizing the axial U(1) identity. Simultaneously, we expect
that the scheme also improves non-anomalous chiral symmetric
properties of the Wilson fermion in QCD. To pursue this program further,
however, we have to treat for example, the fermion self-energy,
for which the Pauli--Villars regularization gives no clue. On this
point, a superimposing of the higher covariant derivative
regularization has been proposed~[\FRO].

We thank S. Kanno for helpful discussions.
The work of H.S.\ is supported in part by the Ministry of
Education, Science, Sports and Culture Grant-in-Aid
for Scientific Research, No.~08240207, No.~08640347 and
No.~07304029.

\refout
\bye